\documentclass[%
pra%
,reprint%
,twocolumn%
,secnumarabic%
,floatfix%
,showpacs%
,amssymb, amsmath, nobibnotes, aps]{revtex4-1}

\usepackage[latin1]{inputenc}
\usepackage{epsfig}
\usepackage{color}

\newcommand{\ind}[1]{_{\text{#1}}}
\newcommand{\bra}[1]{\left<#1\right|}
\newcommand{\ket}[1]{\left|#1\right>}

\begin{document}

\title{Heralded $W$ state preparation using laser-designed super-atoms}

\author{Martin G\"arttner}
\affiliation{JILA, NIST and the University of Colorado, Boulder, Colorado 80309, USA}
\affiliation{Max-Planck-Institut f\"{u}r Kernphysik, Saupfercheckweg 1, 69117 Heidelberg, Germany}

\date{\today}

\begin{abstract}

We propose a scheme for preparing an ensemble of atoms in a maximally entangled $W$ state by exploiting the Rydberg blockade effect. The success of our protocol is indicated by the detection of an ion, which thus serves as a herald for successful entangled state preparation. The fidelity of the preparation scheme is independent of the number of atoms in the ensemble. Therefore, a small cloud of atoms in a dipole trap randomly loaded from a background gas can be reliably prepared in a maximally entangled state despite of atom number fluctuations. 


\end{abstract}

\pacs{67.85.-d, 32.80.Ee, 42.50.Nn, 03.67.Bg}


\maketitle

\section{Introduction}

The ability to reliably prepare multi-particle entangled states is crucial for applications in quantum information science \cite{Horodecki2009} and quantum enhanced metrology \cite{Giovannetti2011}. Dicke states \cite{dicke1954} constitute a particularly robust class of maximally entangled states \cite{Barnea2015}. Their preparation has been demonstrated with photons \cite{Wieczorek2009, Prevedel2009} and ions \cite{haffner2005}. However, these approaches are difficult to scale up to many particles. In experiments with neutral atoms Dicke states have been prepared using spin-changing collisions \cite{luecke2014} or cavity-based quantum non-demolition measurements \cite{Haas2014, McConnell2015}. In the latter case the successful preparation of a Dicke state is heralded by light. In those experiments, the success probability is extremely low if high fidelity is to be achieved. Ensembles of Rydberg atoms with long-range interactions offer a platform that could overcome both the problem of scalability and low success probability. First demonstrations of the preparation of a $W$ state (or first Dicke state) \cite{Duer2000} with Rydberg atoms have been reported \cite{dudin2012c, Ebert2014, Zeiher2015}, however, they face the problem that due to the probabilistic loading of traps the atom number is not know precisely. Since the collective Rabi coupling depends on the atom number this leads to relatively low fidelity of the state preparation. 
We propose a scheme in which the successful preparation of a $W$ state is heralded by the detection of an ion, making use of the possibility perfect Rydberg blockade in small ensembles of atoms. This scheme achieves near-unity preparation fidelity and the atom number uncertainty only deteriorates the success probability, which is still much larger than in experiments with atoms in optical cavities.

\begin{figure}[t]
  \centering
 \includegraphics[width=\columnwidth]{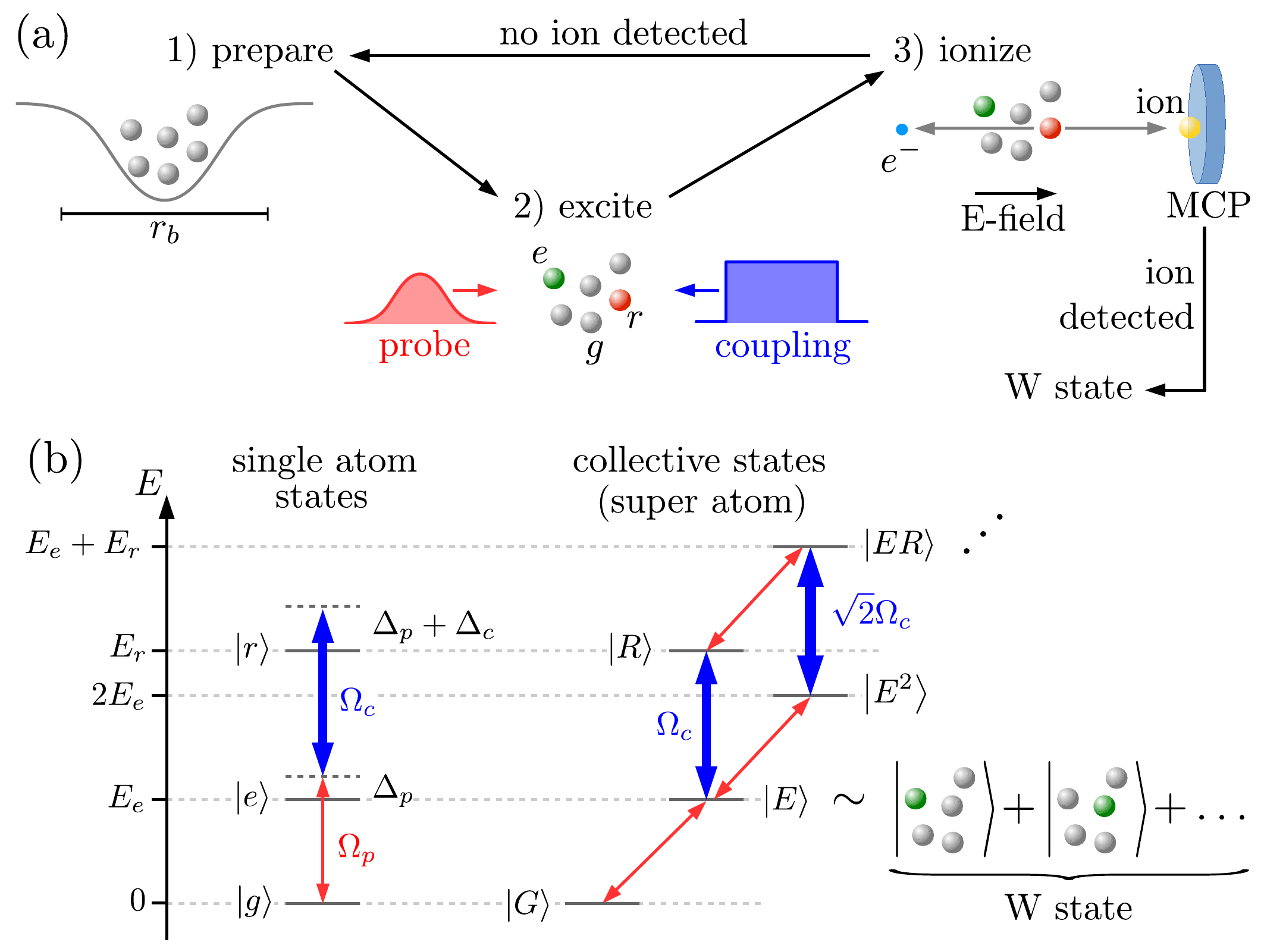}
 \caption{(a) Illustration of the proposed entangled state preparation scheme. The detection of an ion indicates the successful preparation of a $W$ state. (b) Level schemes of single atoms states with laser couplings and symmetrized ensemble states (Dicke states). By strongly coupling pairs of collective states the coupling laser causes AC Stark shifts, which differ between different rungs of the Dicke ladder. This makes it possible to tune the the probe laser to resonance between the collective ground state $\ket{G}$ and only one dressed excited state.}
 \label{fig:scheme}
\end{figure}

The proposed $W$ state preparation scheme is illustrated in Fig.~\ref{fig:scheme}(a). In brief, an ensemble of $N$ atoms is confined to a volume smaller than the Rydberg blockade radius, such that only one atom can be excited to a Rydberg state, and all atoms are initialized in the ground state $\ket{g}$. The ensemble is then excited to a collective state containing one Rydberg excitation $\ket{r}$ and one excitation in a metastable intermediate state $\ket{e}$. Subsequently, the number of Rydberg excitations is measured by field ionization and detection of the removed ion. The remaining $N-1$ particles are then left in the state $\ket{W}=1/\sqrt{N-1}\sum_i \ket{g_1\ldots e_i\ldots g_{N-1}}$. The key point of this scheme is that the successful preparation is \emph{heralded} by the detected ion. The excitation step will not always be successful due to fluctuating atom numbers or other imperfections, but if the Rydberg atom was detected, one can be sure that the $W$ state was prepared successfully. Otherwise the preparation scheme can be repeated until the preparation is successful. 

Essential for this scheme is the ability to drive Rabi-oscillations between two collective states without populating any other states. This is achieved by strongly laser coupling the intermediate and the Rydberg state ($\Omega_c$), as illustrated in Fig.~\ref{fig:scheme}(b), which causes AC-Stark shifts of the collective states. These shifts differ between states with different numbers of excited atoms since the effective Rabi-couplings between these states are collectively enhanced for higher excited states. This allows us to tune the probe field to the resonance between the collective ground state $\ket{G}=\ket{gg\ldots g}$ and a doubly excited state.

\section{Setup and mathematical description}

\begin{figure}[t]
  \centering
 \includegraphics[width=\columnwidth]{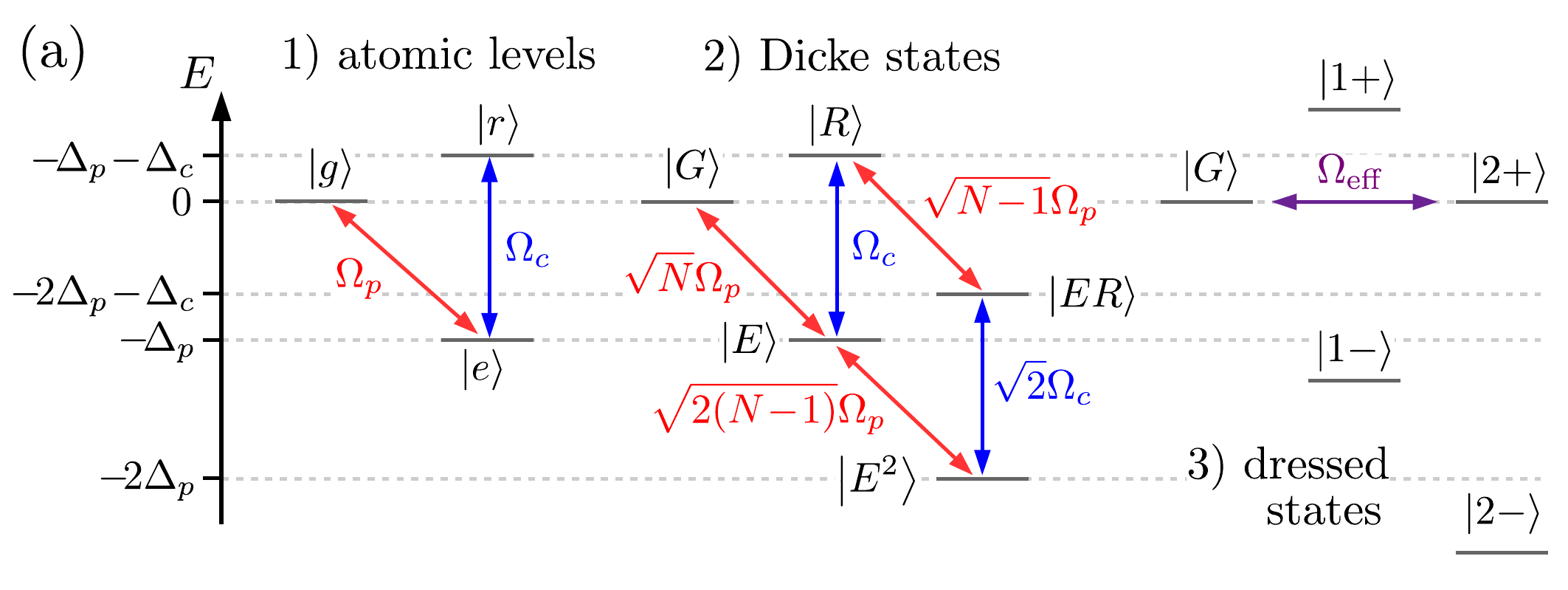}\\
 \includegraphics[width=\columnwidth]{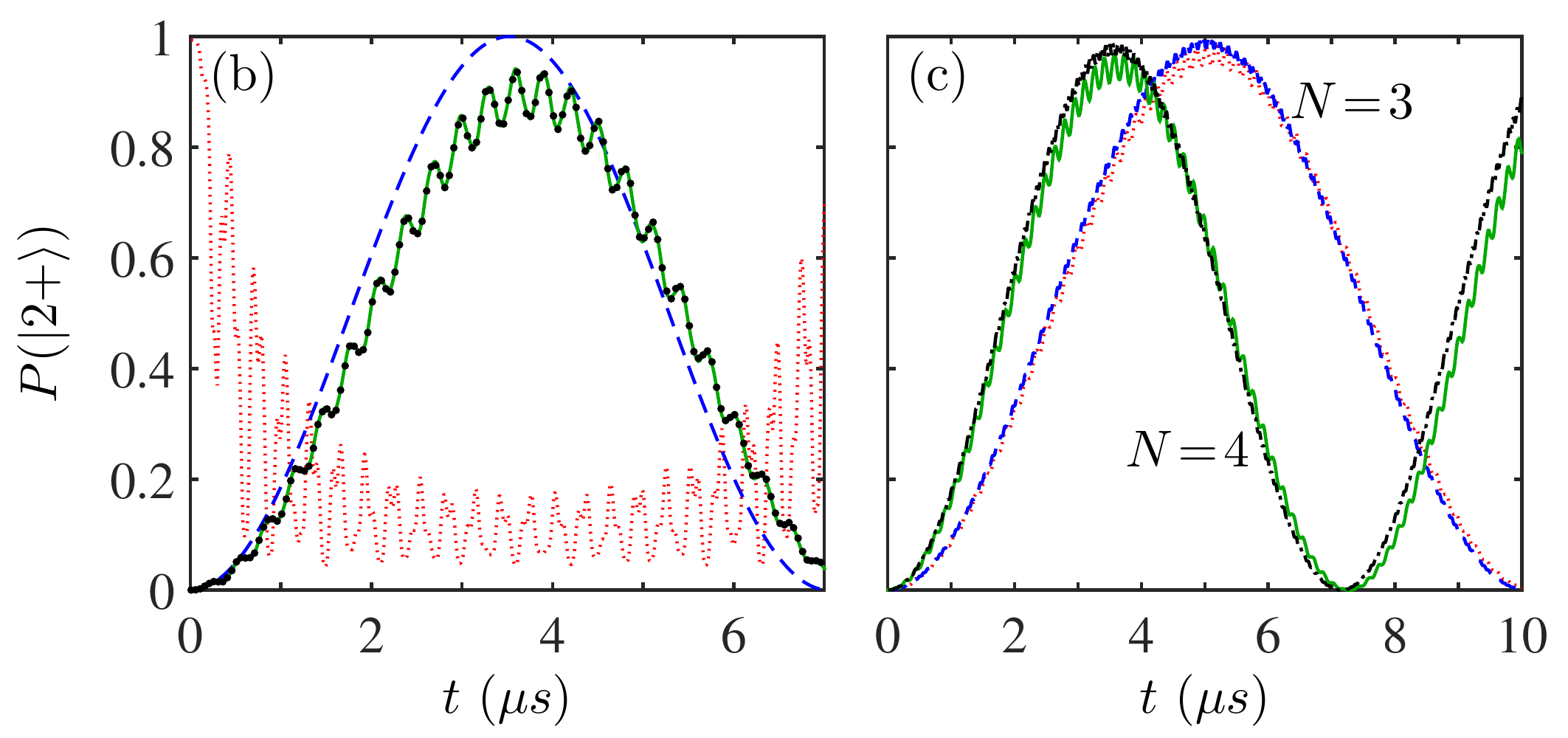}\\
 \caption{(a) Level scheme as in Fig~\ref{fig:scheme}(b), now in the rotating frame. The energy $E$ of the levels are the diagonal elements of the Hamiltonian \eqref{eq:Hamiltonian}. The coupling $\Omega_c$ leads to a repulsion between pairs of Dicke state. We choose the parameters such that the transition between states $\ket{G}$ and $\ket{2+}$ is resonant. (b) and (c) show the Rabi oscillations between these sates. (b) Relatively small $\sqrt{N}\Omega_p/\Omega_c$. Imperfections due to coupling to singly and triply excited sates are visible. We compare solutions of the time dependent Schr\"odinger equation on the full state space (black dots) to the restriction to the lowest dressed Dicke states (solid green line) and the two-state model with all but states $\ket{G}$ and $\ket{2+}$ eliminated (dashed blue line). The red dotted line shows the ratio of Rydberg excitation probability in states other than $\ket{ER}$ over the total Rydberg probability.  Parameters are $N=4$, $\Omega_c/2\pi=10\,$MHz, $\Omega_p/2\pi=0.7\,$MHz, $\Delta_c=-\Omega_c/2$, $\Delta_p=-\Delta_p+\Delta\ind{eff}/2$. (c) Different Rabi frequencies and atom numbers. Solid green and dotted red line: $\Omega_c/2\pi=20\,$MHz, $\Omega_p/2\pi=1\,$MHz. Dot-dashed black and dashed blue line: $\Omega_c/2\pi=50\,$MHz, $\Omega_p/2\pi=1.6\,$MHz. $\Omega_p$ was adjusted to yield the same period of the effective Rabi oscillations for each $\Omega_c$. For small $\sqrt{N}\Omega_p/\Omega_c$ near-perfect transfer to the state $\ket{2+}$ is achieved.}
 \label{fig:Rabi_osc}
\end{figure}

In the following we introduce the mathematical framework employed to simulate the preparation scheme outlined above and to assess its fidelity. The atoms are regarded as motionless on the timescale of the experimental sequence. For now, all three atomic levels are assumed to be non-decaying. These assumptions and concrete experimental realizations will be discussed in Secs.~\ref{sec:issues} and \ref{sec:experiment}. 
The Hamiltonian that governs the dynamics of an ensemble of $N$ atoms, subject to probe- and coupling lasers resonantly driving the transitions between the ground state $\ket{g}$ and the intermediate state $\ket{e}$ and between $\ket{e}$ and the Rydberg state $\ket{r}$, respectively, in rotating wave approximation, reads ($\hbar=1$)    
\begin{equation}
H=\sum_{i=1}^N \left[  H_L^{(i)} - \Delta_p\ket{e_i}\bra{e_i} - (\Delta_p+\Delta_c)\ket{r_i}\bra{r_i}\right]
 \label{eq:Hamiltonian}
\end{equation}
where $H_L^{(i)}=\Omega_p/2 \ket{g_i}\bra{e_i} + \Omega_c/2 \ket{e_i}\bra{r_i} + h.c.$ and $\Delta_p$ and $\Delta_c$ are the detunings of the probe laser from the $ge$-transition and of the coupling laser from the $er$-transition, respectively. The level scheme of a single atom is illustrated in Fig.~\ref{fig:scheme}(b). The same level scheme, now in the rotating reference frame is shown in Fig.~\ref{fig:Rabi_osc}(a), where the given energies are the diagonal elements of the Hamiltonian \eqref{eq:Hamiltonian} and the double arrows symbolize off-diagonal elements, or couplings, between the states. In the given case the probe laser is blue detuned ($\Delta_p>0$) and the coupling laser red detuned ($\Delta_c<0$) with respect to the atomic transition energies.
We assume perfect Rydberg blockade \cite{lukin2001}, which means that the Rydberg-Rydberg interactions are only reflected by the fact that many-body states with more than one atom in the Rydberg state are excluded from the Hilbert space. Thus, in the Hamiltonian itself no interaction term appears. 

The Hamiltonian \eqref{eq:Hamiltonian} is invariant under exchange of particles. Therefore, it is useful to transform the problem to a basis of collective states that also show this symmetry, which is the Dicke state basis. In the absence of decoherence effects the dynamics is restricted to the manifold of symmetric Dicke states, which tremendously simplifies its theoretical treatment \cite{dicke1954, lukin2001, carmele2013, liu2014}.
These states are fully symmetric superpositions of all excited states with the same number of $e$ and $r$ excitations. We label them as
\begin{equation}
 \ket{E^j R^s}=\mathcal{N}_{j,s}\left(\sum_{i=1}^N\ket{e_i}\bra{g_i}\right)^j\left(\sum_{i=1}^N\ket{r_i}\bra{g_i}\right)^s\ket{G}
\end{equation}
with $\mathcal{N}_{j,s}=1/\sqrt{(N-j-s)!/[N^s(N-s)!j!]}$. We use the shorthand notation $\ket{G}\equiv\ket{E^0 R^0}$, $\ket{E}\equiv\ket{E^1 R^0}$ and $\ket{R}\equiv\ket{E^0 R^1}$. Note that the index $s\in\{0,1\}$ for a perfectly blockaded ensemble. The $\Omega_c$ part of the Hamiltonian only couples states with equal $j+s$ which means that in the case of perfect blockade, the coupling laser Hamiltonian splits into two-by-two blocks as illustrated in Fig.~\ref{fig:Rabi_osc}(a2). We further simplify the problem by diagonalizing the coupling laser Hamiltonian and working in the resulting dressed state basis. The resulting dressed state energies (diagonal part of the Hamiltonian in the dressed basis) are shown in Figure~\ref{fig:Rabi_osc}(a3). The key idea of our proposal is now that, by adjusting the parameters $\Delta_p$, $\Delta_c$, and $\Omega_c$, the energy of the state $\ket{2+}$ can be tuned into resonance with the state $\ket{G}$ while all other states are far off-resonant. This choice will result in coherent Rabi oscillations between $\ket{G}$ and $\ket{2+}$, while the populations of all other states stay small.

\section{Dynamics: Collective oscillations}

Figure~\ref{fig:Rabi_osc} illustrates that Rabi oscillations are indeed observed and perfect population transfer from $\ket{G}$ to $\ket{2+}$ is achieved if the parameter $\sqrt{N}\Omega_p/\Omega_c$ is small. Thus, the excitation step of the proposed preparation scheme will be a weak probe pulse that induces complete population inversion ($\pi$-pulse). Since the state $\ket{2+}$ is a superposition of the Dicke states $\ket{E^2}$ and $\ket{ER}$, ideally, the induced dynamics involves only these states and the state $\ket{G}$. This means if a Rydberg atom is detected in the ionization step, the system was in state $\ket{ER}$ previously, and the measurement projects the remaining $N-1$ atoms onto the Dicke state $\ket{E}$, which is the desired $W$ state. As shown in Fig.~\ref{fig:Rabi_osc}(b) the period of the effective Rabi oscillations depends on $N$, which means that the population transfer will in general be imperfect if $N$ fluctuates. However, if the applied laser pulse is not a precise $\pi$-pulse, only the probability of ion detection decreases, but the probability for successful $W$ state preparation \emph{if} an ion was detected, does not. The probability of detecting a Rydberg excitation is what we call the \emph{success probability}, while the probability that in the case of an ion detection a $W$ state was indeed prepared is termed \emph{fidelity}. The fidelity is illustrated by the red dotted line in Fig.~\ref{fig:Rabi_osc}(a), which shows the fraction of detected ions not stemming from the state $\ket{ER}$ (infidelity, fraction of false heralds). This quantity has a wide plateau around $t=\pi/\Omega_{\textrm{eff}}$, which illustrates the tolerance of the scheme with respect to excitation with pulses of imprecise length and thus to fluctuating atom numbers.

What leads to a decrease of the state preparation fidelity is the admixture of states other than $\ket{G}$ and $\ket{2+}$ which contain Rydberg excitations. In order to avoid this, the (AC Stark shifted) dressed state energies of all other states should differ significantly from zero on the level of the typical probe-Rabi-coupling, i.e., $\Omega_c\gg\sqrt{N}\Omega_p$ [see Fig.~\ref{fig:Rabi_osc}(b2)]. In this parameter regime the relevant state space can be restricted to the low excitation sector. It turns out to be sufficient to restrict to the dressed states $\ket{G}$, $\ket{1+}$, $\ket{1-}$, $\ket{2+}$, $\ket{3+}$, and $\ket{3-}$, which is demonstrated in Fig.~\ref{fig:Rabi_osc}(a) by comparing to calculations on the full state space. By adiabatic elimination of all but the states $\ket{G}$ and $\ket{2+}$ we further obtain analytic expressions for effective Rabi-frequency $\Omega\ind{eff}$ and detuning $\Delta\ind{eff}$, which will be given below for a specific set of parameters.

\section{Constraints and parameter optimization.}

In the following we study the performance of the scheme more systematically and optimize the laser parameters. 
The requirement of degeneracy between the states $\ket{G}$ to $\ket{2+}$ leaves a certain freedom in the choice of parameters. For a given coupling strength $\Omega_c$, we can in principle choose the detuning $\Delta_c$ freely. The degeneracy condition can still always be satisfied by adjusting $\Delta_p$. The effective Rabi frequency $\Omega_{\textrm{eff}}$, which determines the duration of the preparation scheme, will have to be chosen such that we operate on a timescale fast compared to typical decoherence times of the system. Choosing a specific $\Omega_{\textrm{eff}}$ will fix the probe Rabi frequency $\Omega_p$. We characterize the quality of our scheme by the two quantities mentioned above: The success probability of the state preparation corresponds to the probability of exciting the target state $\ket{ER}$, and the infidelity, or probability of false heralds, is the probability of Rydberg excitation in states other than $\ket{ER}$ divided by the total Rydberg probability.

\begin{figure}[t]
  \centering
 \includegraphics[width=\columnwidth]{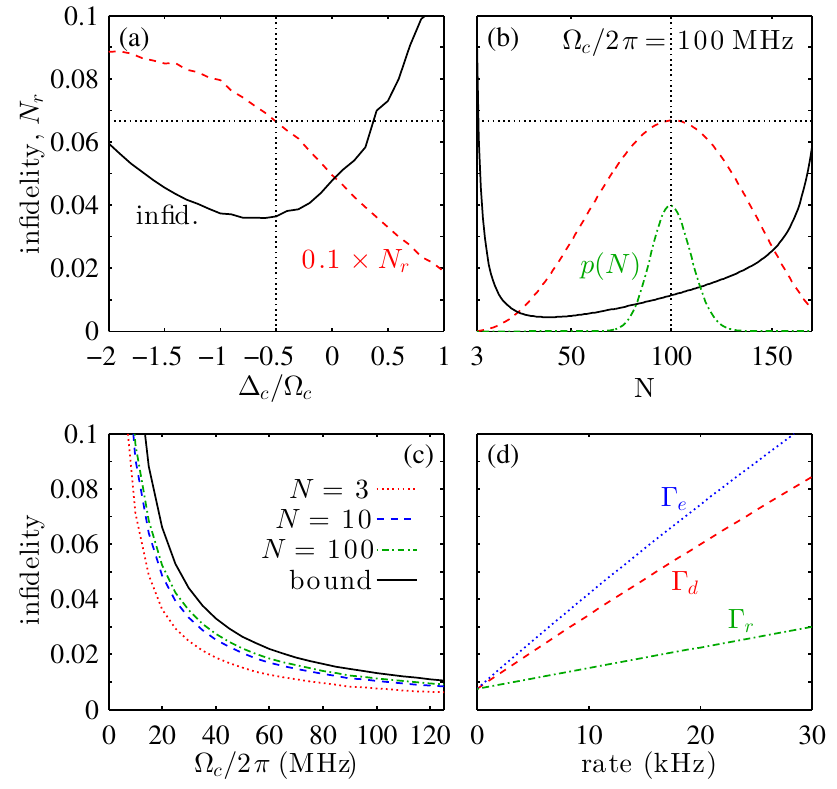}
 \caption{Parameter study of success probability and fidelity of the $W$ state preparation protocol. (a) Rydberg probability $N_r$ (success probability) and fidelity as a function of $\Delta_c$. The fidelity shows an optimum at approximately $\Delta_c=-\Omega_c/2$. The success probability is $2/3$ at the optimal point and approaches 90\% at $\Delta_c/\Omega_c=-2$. Parameters: $N=3$, $\Omega_c/2\pi=20\,$MHz. Here, as in (c) and (d), $\Omega_p$ was adjusted such that $\pi/\Omega\ind{eff}=5\,\mu$s and $\Delta_p$ is determined by the resonance condition. In panels (b)-(d) the condition $\Delta_c=-\Omega_c/2$ was imposed. (b) Dependence on the atom number $N$ for fixed laser parameters. The laser parameters are chosen such that the success probability is maximal for $N=100$ atoms. The dot-dashed green line shows the probability $p(N)$ to prepare an ensemble of $N$ atoms for a Poissonian distribution with a mean of $N=100$ atoms. The protocol shows reasonable success probability and fidelity across the whole atom number distribution. (c) Fidelity as a function of $\Omega_c$ for different atom numbers. At $\Omega_c/2\pi>100\,$MHz, fidelities beyond 99\% can be achieved. The black line shows an estimated upper bound on the fidelity. (d) Dependence on different sources of decoherence. The fidelity increases linearly with spontaneous decay rate from intermediate ($\Gamma_e$) and Rydberg state ($\Gamma_r$), as well as with the single-atom dephasing rate $\Gamma_d$.}
 \label{fig:infidelity}
\end{figure}

Figure~\ref{fig:infidelity}(a) shows these two quantities as a function of the chosen detuning of the coupling laser $\Delta_c$. We used $\Omega_{\textrm{eff}}/2\pi=0.1\,$MHz, corresponding to an excitation time of $\pi/\Omega\ind{eff}=5\,\mu$s. We find that the minimal infidelity is reached close to $\Delta_c=-\Omega_c/2$, marked by the dotted vertical line. Here, the probability of preparing the state $\ket{2+}$ is approximately $2/3$. Note that the qualitative behavior, especially the position of the minimum infidelity is largely independent of the choice of $N$ and $\Omega_c$. In the following we focus on the choice $\Delta_c=-\Omega_c/2$, for which the degeneracy condition $E(\ket{2+})=0$ implies $\Delta_p=-\Delta_c$. For this choice one obtains the dressed state $\ket{2+}=(|E^2\rangle+\sqrt{2}\ket{ER})/\sqrt{3}$. The adiabatic elimination yields the effective Rabi frequency $\Omega_{\textrm{eff}}=\sqrt{2/3}\sqrt{N(N-1)}\Omega_p^2/\Omega_c$ and the effective detuning $\Delta_{\textrm{eff}}=(2 N-7)/3\, \Omega_p^2/\Omega_c$. The effective detuning can be compensated (approximately) by modifying the probe detuning to $\Delta_p=-\Delta_c+\Delta_{\textrm{eff}}/2$ (as also done in Fig.~\ref{fig:Rabi_osc}).

In order to demonstrate that this fidelity is not decreased much by fluctuations in the atom number $N$ we calculated the mean fidelity for a Poissonian atom number distribution with mean $\langle N\rangle=100$ for $\Omega_c/2\pi=100\,$MHz. The result is shown in Fig.~\ref{fig:infidelity}(b). The weighted average over the Poisson distribution gives an infidelity of $1.15\%$ instead of $1.14\%$ obtained for $N=100$ with an optimal $\pi$-pulse. This also illustrates that the excitation time can be chosen much smaller than $\pi$-pulse without loss of fidelity, but only at the expense of a smaller success probability, which makes the scheme robust not only with respect to  atom number fluctuations but also to imperfect control of the spatial and temporal shape of the probe laser pulse. 

In Fig.~\ref{fig:infidelity}(c) we show the infidelity as a function of $\Omega_c$ for different atoms numbers keeping $\Omega_{\textrm{eff}}/2\pi=0.1\,$MHz fixed. The infidelity decreases as $\Omega_c^{-1}$ and becomes smaller than $1\%$ for Rabi frequencies exceeding $100\,$MHz. By estimating the off-resonant admixture of states containing Rydberg excitations other than $\ket{ER}$ we find that $10\Omega_{\textrm{eff}}/\Omega_c\propto N\Omega_p^2/\Omega_c^2$ presents an upper bound on the infidelity.

\section{Possible issues}
\label{sec:issues}

So far we have assumed perfectly coherent dynamics. However, in a realistic scenario various sources of decoherence, such as spontaneous emission, finite laser line-widths, and atomic motion will be present. In the following we discuss the effect of such imperfections on the fidelity of our protocol. 
In order to include decoherence effects the system is described by a density operator obeying the master equation (ME)
\begin{equation}
 \dot{\rho}=-i[H,\rho]+\mathcal{L}[\rho] \,.
\end{equation}
Decoherence effects can be included via Lindblad terms 
\begin{equation}
 \mathcal{L}_k[\rho]=\Gamma_k\left(s_{k}\rho s_{k}^\dagger - 1/2\left\{s_{k}^\dagger s_{k},\rho\right\}\right)
\end{equation}
where, e.g., $s_{k}=\ket{g_k}\bra{e_k}$ ($s_{k}=\ket{e_k}\bra{r_k}$) for spontaneous emission from the intermediate (Rydberg) level, respectively, and $s_{k}=\ket{r_k}\bra{r_k}$ for single-atom  dephasing of the Rydberg state ($k$ is the atom index). To account for dephasing effects that do not affect all atoms independently but the ensemble as a whole, we consider projectors on the collective states $S\ind{coll}=\ket{E^jR^s}\bra{E^jR^s}$.

Decay of the intermediate level leads to incoherent population transfer from $\ket{ER}$ to singly excited states involving the unwanted state $\ket{R}$ and thus decreases the fidelity. Decay from the Rydberg state will harm the proposed scheme less, since such a decay leaves the system in a state without Rydberg excitation and thus does not produce false heralds. The infidelity as a function of decay rates is shown in Fig.~\ref{fig:infidelity}(d) for $N=3$ atoms, $\Omega_c/2\pi=100\,$MHz, and $\Omega_{\textrm{eff}}/2\pi=0.1\,$MHz. The infidelity increases linearly with the time-integrated decay probability ($\propto\pi\Gamma_e/\Omega_{\textrm{eff}}$), where the proportionality factor is about a factor of five smaller for the dependence on $\Gamma_r$ compared to $\Gamma_e$.

Other decoherence effects include finite laser linewidths, atomic motion, and stray electric fields Stark shifting the Rydberg state. These effects can result in single-atom or collective dephasing. Collective dephasing here means, that the coherences between the symmetric Dicke state are damped but no coupling to asymmetric states is caused as the perturbing process also conserves the exchange symmetry of the atoms. Such process do not harm our scheme much.
Single atom dephasing, on the other hand, leads to the distribution of population into states outside the symmetric Dicke state manifold and therefore reduces the probability to prepare the symmetric state $\ket{ER}$. Again, the infidelity increases linearly with the dephasing rate as shown in Fig.~\ref{fig:infidelity}(d).

To avoid imperfection due to decoherence short excitation times (large $\Omega\ind{eff}$) are required. On the other hand we have seen that in order to minimize couplings to unwanted state small ratios of $\Omega\ind{eff}/\Omega_c$ are favorable. Thus, in order to reach the maximum fidelity of $W$ state preparation for given $\Omega_c$ and given rates of the decoherence process, the effective Rabi frequency (or duration of the protocol) can be optimized and the maximum fidelity for given atomic parameters depends crucially on the achievable coupling laser strength. 

Another possible issue is whether the ion and electron that might traverse the atomic ensemble after ionization would degrade the fidelity by inducing local Stark shifts or collisions with the remaining atoms. 
In order to assess this effect we performed classical Monte Carlo trajectory simulations assuming that the ionizing electric field is ramped from $0$ to $1\,$kV/cm in $300\,$ns \cite{hofmann2013}. As an example we use the differential $dc$-polarizability $\Delta\alpha$ of $^{88}$Sr clock transition \cite{middelmann2012} to estimate the induced local phase shifts. The ion leaves a trap of diameter $1\,\mu$m in about $25\,$ns. For a trap of volume $1\,\mu$m$^3$ containing $N=100$ atoms, the probability to induce a significant phase shift in an atom-ion collision turns out to be smaller than $10^{-4}$. The ionizing electric field itself also does not affect the coherence between states $\ket{g}$ and $\ket{e}$ significantly.

The proposed scheme assumes that one operates in the regime of perfect Rydberg blockade, meaning that the interaction shift is much larger than the laser line widths and coupling strengths $\hbar \Omega$. This means that the ensemble size should not exceed a few micrometers. At the same time the mean distance between the particles must not become too small such that effects of the interaction of the Rydberg electron with the surrounding ground state atoms can be neglected. This limit the atom numbers that can be used to a few tens or hundreds. A more detailed account of the effects of imperfect blockade and ground-Rydberg state interactions are left for future studies.

\section{Experimental implementation}
\label{sec:experiment}

We have seen in the previous section that the most important requirement for achieving high fidelities in the proposed scheme is a small spontaneous decay rate from the intermediate level $\ket{e}$. Note that using, e.g., the $5P_{3/2}$ state of Rubidium as intermediate state is not possible due to its fast decay rate of tens of MHz. Two possible ways to overcome this problem are outlined in the following.

Using Rubidium or other alkali atoms, one possibility is to use two hyperfine ground states as $\ket{g}$ and $\ket{e}$. The coupling to the Rydberg state can then either be achieved by a two-photon transition via an (off-resonant) optically excited state or by directly coupling to Rydberg state using a laser in the UV-range \cite{hankin2014, manthey2014, Sassmannshausen2015}. The transition between the two lower states can either be driven by microwaves or by a two-photon Raman transition. Such schemes have recently been successfully implemented experimentally \cite{isenhower2010, hankin2014}. The Raman scheme might also allow for a Doppler free implementation \cite{Keating2015}.

A second possiblity is to use alkaline earth (like) atoms, such as Ca, Mg, Sr, or Yb, that are promising because of their metastable triplet $P$-states, which have been extensively used in optical atomic clocks (see Ref.~\cite{poli2013} for a recent review). Many alkaline earth (like) species have been laser-cooled to microkelvin temperatures and their transitions from the $^1S_0$ ground state to states $^3P_1$ or even the doubly forbidden $^3P_0$ offer a wide range of linewidths (see e.g. Table V in Ref.~\cite{poli2013}).
From the intermediate state, $5sns^3S_1$ Rydberg states can be excited by blue or near UV-lasers. The Rydberg interaction potentials are similar to those of alkaline atoms for s-states (near-isotropic van der Waals) \cite{gil2014}. Atomic structure calculations for these states have recently been reported \cite{vaillant2014, Vaillant2012}. Also, there have been first experiments using Rydberg states of $^{88}$Sr \cite{millen2010, mcquillen2013, lochead2013}. 

The greatest challenge for both implementations will be to build strong enough blue lasers. Here, recent efforts in the context of Rydberg dressing and direct Rydberg excitation from the ground state could lead to significant advances in the near future \cite{Weber2015, hankin2014}. For example, tunable, solid-state lasers producing $\approx0.5 W$ of narrow-band cw light have been developed for this wavelength regime \cite{wilson2011, friedenauer2006, vasilyev2011}. 

We also note that most experiments the trapping potential would be anti-trapping for the Rydberg states and thus has to be switched off during the preparation sequence involving the population of Rydberg states. For a further processing of the prepared $W$ states it might be necessary to recapture the ensemble after the preparation, which may pose an additional experimental challenge. However, if recapture or operation in a trap with magic wavelength Rydberg trapping could be achieved, this would allow to repeat the preparation sequence [Fig.~\ref{fig:scheme}(a)] without having to reload the trap in every cycle but by re-initializing the internal state of the atoms by optical pumping into the ground state. This might reduce the atom number in the sample in cases where a Rydberg atom was present but was not detected due to the finite efficiency of ion detection. Note, however, that current experiments reach 90\% detection efficiency \cite{Teixeira2015}.

\section{Conclusions and outlook}

In conclusion we have proposed a scheme for the preparation of a maximally entangled $W$ state heralded by the detection of an ion. This scheme could be implemented using small clouds of ultracold alkali or alkaline earth atoms in a small optical dipole trap. 

There are several possible extensions of this idea, which one could think of.

So far, we restricted ourselves to the low $j$ part of level structure of the artificial atom created by the coupling laser. Higher excited collective states could be prepared using a sequence of probe pulses with different frequencies or by employing adiabatic preparation schemes \cite{petrosyan2013b, noguchi2012}. Due to their multi-particle entanglement, higher excited Dicke states have many applications in metrology \cite{krischek2011, toth2012, hyllus2012} and quantum information science \cite{chiuri2012, Pal2015}. Dicke states are not spin squeezed in the conventional sense \cite{kitagawa1993, wineland1992}, but nevertheless their entanglement is useful for sub-shot-noise interferomentry. In general, the maximum fringe contrast that can be obtained is a spectroscopic measurement with a given input state is given by $1/\sqrt{F_Q}$, where $F_Q$ is the quantum Fisher information \cite{ma2011, toth2009, hyllus2010}. 
For Dicke states of $N$ spins with $m$ excitations, one obtains $F_Q=N+2m(N-m)$, so for a $W$ state $(m=1)$ we have $F_Q=3N-2$, which means that the fringe contrast can be enhanced at most by a factor $\sqrt{3}$ compared to the shot-noise limit. Thus, ideally one would like to prepare the state with $m=N/2$, corresponding to $\ket{E^{N/2}}$ in the notation used above. This state enables spectroscopy at the Heisenberg limit \cite{luecke2014} and is also useful for quantum information processing tasks, such as $1\rightarrow (N-1)$ telecloning or open-destination teleportation \cite{chiuri2012}.

Furthermore, the Hamiltonian of the artificial atom can be mapped to the Jaynes Cummings model, where the role of the $e$-excitation is taken by the photons and the $r$-state corresponds to the state of the two-level atom in the cavity. In the Dicke basis the Hamiltonian reads
\begin{equation}
\begin{split}
 H=H_p+&\frac{\Omega_c}{2}\sum_{j=0}^{N-1}(\sqrt{j+1}\ket{E^{j+1}}\bra{E^j R}+h.c.) \\
 &- \Delta_c\sum_{j=0}^{N-1}\ket{E^j R}\bra{E^j R}
\end{split}
\end{equation}
Thus, the probe field $H_p$ would account for a coherent external drive of the cavity while the coupling laser part resembles the Hamiltonian of a single atom in a cavity, where $j$ is the number of cavity photons and the atomic ground (excited) state corresponds to the absence (presence) of a Rydberg excitation. One could envision an experiment, in which first a probe pulse is applied which creates a coherent superposition of the states $\ket{E^j}$ (binomial distribution). Then switching on the coupling laser would induce collapse and revival dynamics of the Rydberg population, characteristic for the Jaynes-Cummings model.

Another interesting question concerns the motional state of the ensembles after the removal of the Rydberg atom. Can the ionization be done in a coherent way, such that a superposition of different position states is created instead of an incoherent mixture? For example, in a setup  with high control of atomic positions as in \cite{isenhower2010, hankin2014, Barredo2014, Nogrette2014, Zeiher2015} states entangled not only in the internal, but also in the external degrees of freedom might be prepared that way.

\begin{acknowledgments}
We thank
A.\ M.\ Rey, 
J.\ Evers, and S.\ Whitlock
for helpful discussions. 
\end{acknowledgments}

\end{document}